\documentclass[twocolumn,aps,floatfix,nofootinbib]{revtex4}
\usepackage{latexsym}
\usepackage{mathptm}
\usepackage{amsmath}
\usepackage{amssymb}
\usepackage{graphicx}

\newcommand{\f}{\begin{equation}}
\newcommand{\ff}{\end{equation}}
\newcommand{\be}{\begin{equation}}
\newcommand{\ee}{\end{equation}}
\newcommand{\barray}{\begin{array}}
\newcommand{\earray}{\end{array}}
\newcommand{\bea}{\begin{eqnarray}}
\newcommand{\eea}{\end{eqnarray}}
\newcommand{\rd}{\mathrm{d}}

\begin{document}

\title{  Relative locality and the soccer ball problem  }
\author{Giovanni Amelino-Camelia$^a$, Laurent Freidel$^b$, Jerzy Kowalski-Glikman$^c$, Lee Smolin$^b$\thanks{lsmolin@perimeterinstitute.ca}
\\
$^a$Dipartimento di Fisica, Universit\`a La Sapienza and Sez.~Roma1 INFN, P.le Moro 2, 00185 Roma, Italy\\
$^b$Perimeter Institute for Theoretical Physics, 31 Caroline
Street North, Waterloo, Ontario N2J 2Y5, Canada\\
$^c$Institute for Theoretical Physics, University of Wroclaw,  Pl. Maxa Borna 9, 50-204 Wroclaw, Poland }

\date{\today}

\begin{abstract}

We consider the behavior of macroscopic bodies within the framework of relative locality \cite{PRL}, which is a recent proposal
for Planck scale modifications of the relativistic dynamics of particles which are described as arising from deformations
in the geometry of momentum space.  These lead to the addition of non-linear terms to the energy-momentum relations
and conservation laws, which are suppressed by powers of ratio between the energy $E$
of the particles involved and the Planck mass $M_P$.
We consider and resolve a common objection against such
proposals, which is that, even if the corrections are small for elementary
particles in current experiments, they are huge when applied to
composite systems such as soccer balls, planets and stars,
with energies $E_{macro}$ much larger than $M_{P}$.
We show that
this {\it "soccer-ball problem"} does not arise within the
framework of  { relative locality}, because the non-linear effects
for the dynamics of a composite system with $N$ elementary particles
appear at most of order  $E_{macro}/ N \cdot M_{P}$.

\end{abstract}

\maketitle


\section{Introduction}

There are general reasons to suspect the existence of a regime of quantum gravity phenomena
which may manifest itself as corrections to the basic relations of special relativistic particle dynamics
of order of powers of $\mbox{Energies}/M_P$ where the Planck mass is $M_{P} = \sqrt{\frac{\hbar}{G_{Newton}}}$.
Over the last decade several different experiments and astrophysical observations have
reached sensitivity levels suitable for testing the presence
of such terms, making this regime a possible
site for the first experimental discovery of quantum gravitational phenomena.
In a recent paper, \cite{PRL} we proposed a general framework called {\it relative locality} which encompasses a class of such theories, based on the
notion that the deformations of special relativistic physics can be coded by curvature and other non-linear deformations of
momentum space.

In these theories we expect that the energy-momentum relations of relativistic particles may receive non-linear
corrections which arise from attributing to momentum space,  $\cal P$,  a non-trivial curved metric.  The energy-momentum relations
then take the general form
$
D^{2}(p) = m^{2}.
$
Here $D(p)$ is interpreted as the {\it distance} from $p$ to the zero energy momenta $0$ with respect to
a Lorentzian metric $ g^{\mu\nu}(p)\rd p_{\mu}\rd p_{\nu}$ on  momenta space.  To see how these lead to corrections
to special relativity, imagine that we choose coordinates on momentum space, so that the metric can be written
as the Minkowski metric plus terms of order $\frac{E}{M_P}$.  Then we will have
\begin{equation}\label{1}
   D^{2}(p) = E^2-  \vec{p}^2 -  \eta \frac{E}{M_P} \vec{p}^2+ \cdots = m^2
\end{equation}
with $\eta$ being a numerical coefficient. We will shortly discuss how one chooses such coordinates, but for the
moment let us proceed naively as this will suffice to state the problem this letter solves.

Similarly, non-linear corrections to the conservation laws of energy and momentum can be understood as arising from
the choice of a non-trivial connection on momentum space, $\cal P$.  As shown in \cite{PRL} these arise from a new, non-linear
addition rule for momentum.
In the simplest case of the
two-to-one process $A + B \rightarrow C$,  this takes the form
\begin{equation}
\label{2}
    p^{(C)}_\mu = \left(p^{(A)} \oplus p^{(B)}\right)_\mu\, ,
\end{equation}
We assume again the existence of a nice set of coordinates on $\cal P$ which allows us to expand this as
\begin{equation}\label{3a}
\left(p^{(A)} \oplus p^{(B)}\right)_\mu =p^{(A)}_\mu + p^{(B)}_\mu -
\frac{1}{M_P}  \tilde{\Gamma}_\mu{}^{\alpha\beta}\, p^{(A)}_\alpha\,
p^{(B)}_\beta+ \cdots
\end{equation}
Here and in the similar formulas below ${\tilde \Gamma}$ denotes the
($M_p$-rescaled) connection coefficients on momentum space evaluated
at the origin $p_{\mu}=0$. Notice that the
 connection coefficients
 on momentum space $\Gamma_\mu{}^{\alpha\beta}(0)$
($\equiv \tilde{\Gamma}_\mu{}^{\alpha\beta}/M_p$)
have dimensions of inverse mass, so the ${\tilde \Gamma}$ are dimensionless.

In both cases we write down only the leading order corrections to the
standard special relativistic expressions. This is justified in the
case of elementary particles, because even for most energetic cosmic
rays the ratio of their energies to $M_P$ is of order of $10^{-8}-10^{-9}$,
and the higher order terms can be safely neglected.

\section{Statement of the soccer ball problem}

We can now state the {\it soccer ball problem} which this letter is
addressed to. The laws of special relativistic dynamics are
universal.  They apply equally to elementary particles and to large
macroscopic bodies such as planets, stars and soccer balls. Either
this is the case also with relative locality or it is not the case.
If it is the case that the same laws apply to soccer balls and other
macroscopic bodies, then the proposal is clearly wrong, the
corrections are of order ${M_{\mbox{soccer ball}}} / {M_P}$.  Since
the Planck mass is roughly of the order of $10^{-5} g$ this is a
huge quantity;  while we do  experience to very  good accuracy that
macroscopical objects  follows a linear addition of momenta.
  This is the soccer ball problem: A serious objection often
  raised (see, {\it e.g.}, Refs.~\cite{maggioNOsoccer,sabiNOsoccer})
  to dismiss any attempt to introduce non linearities in momentum space.

  Relative locality, as formulated in  \cite {PRL}, is positing that the fundamental equations of particle dynamics apply
only to elementary particles. This is a departure from previous
theories in which the laws of motion are universal, and is perhaps
necessary to unify gravity with quantum theory. Then we still have
to ask, what laws are nonetheless implied for the dynamics of
soccer-balls, planets and stars?  Can we derive these laws and show
that there are no observable deviations from the predictions of
special relativity.

To answer this query we proceed in two steps.  First we will give a simple,
but rather naive, answer.  This serves only to highlight the key point.
Then we give a more rigorous  argument.  The difference between the naive and the more rigorous argument, as we will see, mainly comes from being
careful about the choice of coordinates on momentum space made in the course of the argument.  That some choice is needed is clear,
because the possibility of expanding the metric in (\ref{1}) around the flat metric plus terms of order $\mbox{E}/M_P$ depends on the choice of
coordinates.

\section{A naive argument}

Here is the naive argument.   Let us consider first the  modified dispersion relation (\ref{1}).  The key observation
is that the soccer ball is not an elementary system, it is composed
of a huge number $N$ of elementary particles. Let each elementary
particle be described by mass shell relation (\ref{1}) and assume,
for simplicity, that all the particles have identical masses $m$ and
momenta $p_\mu$. The total mass of the ball is therefore
$M_{(ball)}= N\, m$ and its total momentum is $\mathbf{P}_{(ball)}{}_\mu =
N\, p_\mu$. Substituting this to (\ref{1}) we easily see that
\begin{equation}\label{4}
    E_{(ball)}^2=  \vec{\mathbf{P}}_{(ball)}^2 + M_{(ball)}^2 + \eta\, \frac{E_{(ball)}}{N M_P} \vec{\mathbf{P}}_{(ball)}^2+ \ldots
\end{equation}
Comparing (\ref{1}) with (\ref{4}) we see that although the
deformation is still present its magnitude is governed now not by
the scale $M_P$ but by the scale $N$ times bigger, which renders the
last term in (\ref{4}) negligible for all practical purposes.

Let us apply the same naive argument to the non-linear conservation law (\ref{3a}).
Let us now assume that we deal with two macroscopic bodies, body $A$ containing $N$ particles
with identical
momenta $p^A _\mu$ and body $B$ also containing $N$ particles with identical momenta $p^B _\mu$.
The total momentum of body $A$ may be, naively, {\it defined} to be just
\f
\mathbf{P}^{(A)}_\mu  = N p^A _\mu
\label{naivedef}
\ff
and likewise for $\mathbf{P}^{(B)}_\mu  = N p^B _\mu$. Let us assume that the collision of the two bodies can be described as
the set of events in which one particle from body $A$ interacts according to (\ref{2}) with one particle from body $B$.
We then easily see that
\begin{equation}\label{3naive}
\left(\mathbf{P}^{(A)}  \oplus \mathbf{P}^{(B)} \right)_\mu =\mathbf{P}^{(A)}_\mu +\mathbf{P}^{(B)}_\mu -
\frac{1}{N M_P}  \tilde{\Gamma}_\mu{}^{\alpha\beta}\, \mathbf{P}^{(A)}_\alpha\,
\mathbf{P}^{(B)}_\beta+ \cdots
\end{equation}
Hence the same conclusion holds that the non-linearities are damped by powers of $N M_P$.

This argument is too naive on two counts.  First, we need to
be specific about the choice of coordinates on momentum space.
Second, the definition (\ref{naivedef}) is not fully justified. We
have made the other equations of relativistic dynamics non-linear;
why not that one also? Shouldn't we expect that the definition of
the total momenta also involves non linearity?    We now address
each in turn and the result is a more rigorous argument to the same
conclusion.

\section{A more rigorous argument}

\subsection{The choice of coordinates on momentum space}

As we stressed,  in order to write down the formulas
(\ref{1}), (\ref{2}),   one has to choose a coordinate system on the
momentum space. The coordinates must be such that the origin
corresponds to the state with zero momentum and that both modified
dispersion relation and momentum composition rules become the
standard special relativistic ones in the limit of vanishing
momentum space curvature or $M_P \rightarrow \infty$. The archetypal
example of such coordinate system is provided by Riemann normal coordinates.
In Riemann normal coordinates the metric geodesics from the origin are straight lines
and we find
\begin{equation}\label{DR}
    m^2 = D^2(p) \equiv \eta^{\mu\nu}p_{\mu}p_{\nu}\, ,
\end{equation}
and therefore the dispersion relation in normal coordinates is not
modified. In this case therefore the whole information about the
momentum space curvature is contained in the deformed momentum
composition rule (\ref{2}). In any  other coordinate
system,  the dispersion relation, still defined
as $m^2=D^2(p)$ would take the general form (\ref{1}).

Another important coordinate which we will use is the connection
normal coordinates, for which the geodesics associates with  the
connection are straight lines, even if the connection is not
metrical. In these coordinates $\hat{p}$ the addition of parallel
momenta is linear i.e. \be (a\hat{p}) \hat{\oplus} (b \hat{p}) =
(a+b) \hat{p} \ee where $a,b$ are any scalars. At first order in
$M_{P}$  the connection coordinates are given by
$\hat{p}_{\mu}=F_{\mu}(p)$ where \be F_{\mu}({p}) = p_{\mu} +
\frac1{2M_{P}} \tilde{\Gamma}_{\mu}^{\alpha \beta} {p}_{\alpha}
{p}_{\beta}+\cdots \ee The addition in the new coordinates is given
by $\hat{p}\hat{\oplus} \hat{q} \equiv F( F^{-1}(\hat{p}) \oplus
F^{-1}(q))$ while its expansion is \be \hat{p}\hat{\oplus} \hat{q} =
\hat{p}_\mu + \hat{q}_\mu - \frac{1}{M_P}\,
\tilde{\Gamma}_\mu{}^{[\alpha\beta]}\, \hat{p}_\alpha\,
\hat{q}_\beta+ \cdots \ee where the bracket denotes
antisymmetrization. Since only the torsion component at $p_{\mu}=0$
enters at first order we obtain the desired result. If the
connection is metrical, that is if $\nabla^{\mu}g^{\alpha\beta}=0$
then the Riemann and connection normal coordinates agree. If the
connection is non metrical the metric geodesics and connection
geodesics no longer agree.  We note that this would have interesting
phenomenological consequences \cite{GRB-ll}.

\subsection{A model of macroscopic bodies in collision}

Now that we have discussed the choice of coordinates we are ready to
present a careful analysis of the soccer ball problem. This involves
an idealization of the properties of macroscopic body, then we will
make it slightly less idealized.

We first consider an  idealized situation, involving two
bodies ``$A$" and ``$B$" each composed of $N$ atoms. Let us
assume that in the course of their interaction the bodies exchange
photons. Denoting the photon's momentum by $k_\mu$ and the initial
and final momentum of the atom by $p_\mu$ and $\tilde p_\mu$,
respectively we find that for the photon emission process we have
\begin{equation}\label{5}
    p_\mu = ({\tilde p} \oplus k)_\mu
\end{equation}
while for photon absorption
\begin{equation}\label{6}
    (k \oplus p)_\mu = {\tilde p}_\mu
\end{equation}
Let us now consider the process of a single photon exchange between
the body $A$ and $B$. Assuming that the body $A$ emits the photon,
and body $B$ absorbs it we find the relations
$$
p_{\mu}^{A} = ({\tilde p}^A \oplus k)_\mu\, , \quad (k \oplus
p^{B})_\mu = {\tilde p}_{\mu}^{B}\, .
$$
Solving these two equations for $k$ we find the
relation.
\begin{equation}\label{7}
    [(\ominus {\tilde p}^A \oplus p^A) \oplus p^{B}]_\mu = {\tilde p}_{\mu}^{B}
\end{equation}
where we introduced the antipode of momentum $\ominus p$ defined by
$(\ominus p) \oplus p =0$ and used the left inverse property $(\ominus p)\oplus (p\oplus q)=q$. In the leading order the antipode is given by
$$
(\ominus p)_\mu = - p_\mu -\frac{1}{M_P}\,
\tilde{\Gamma}_\mu{}^{\alpha\beta}\, p_\alpha\, p_\beta+ \ldots
$$
Eq.\ (\ref{7}) describes the momentum conservation rule of a single
interaction (emission and absorption) process. The soccer ball
problem would arise if the same form of the conservation rule would
hold  for macroscopic, massive bodies with initial and final total
momenta $P^{A,B}_\mu$ and $\tilde P^{A,B}_\mu$, respectively,
i.e. if we had
\begin{equation}\label{8}
    [(\ominus {\tilde P}^A \oplus P^A) \oplus P^{B}]_\mu = {\tilde
    P}_{\mu}^{B}\, .
\end{equation}

Let us now show that this naive expectation is not fulfilled in the
case of a large body composed of a large number of microscopic
subsystems. To see this let us assume that in the interaction
process  each of the $N$ atoms of the body $A$ emits one and only
one photon that is subsequently  absorbed by one and only one
 atoms of the body $B$. For each such process we have to do with
 the conservation rule (\ref{7}) so that for
 each interacting pair of constituents of the bodies $A$ and $B$, labeled by index
 $a$, $a=1,\ldots, N$ we can write
 \begin{equation}\label{9}
[(\ominus {\tilde p}^{A_{a}} \oplus p^{A_{a}}) \oplus p^{B_{a}}]_\mu =
{\tilde
 p}^{B_{a}}{}_\mu
 \end{equation}
Expanding this expression to the leading order in $1/M_P$ we get
the relation
\begin{equation}\label{10}
\begin{split}
 &  [p^{A_{a}} +p^{B_{a}}]_\mu
-\frac1{M_P} \tilde{\Gamma}_\mu^{\alpha \beta} p_{\alpha}^{A_{a}} p_{\beta}^{B_{a}} + \cdots
\\
=& [\tilde{p}^{A_{a}} + \tilde{p}^{B_{a}}]_\mu
-\frac1{M_P} \tilde{\Gamma}_\mu^{\alpha \beta} \tilde{p}_{\alpha}^{A_{a}} \tilde{p}_{\beta}^{B_{a}}
+\cdots
\end{split}
\end{equation}
Note that to leading order the LHS of this equation is just $p^{A_{a}}\oplus p^{B_{a}}$  and the equation expresses that this is equal
 to this order to
$\tilde{p}^{A_{a}} \oplus \tilde{p}^{B_{a}}$. This is a momentum conservation equation which expresses that the non linear addition is the 
one preserved in interactions mediated by photons exchange.

\subsection{The definition of the total momentum of a body}

Let us now define the macroscopic momentum
to be the non linear composition of the microscopic momenta in some order
\begin{equation}\label{11}
    \mathbf{P}^{A} \equiv p^{A_{1}} \oplus (p^{A_{2}}\oplus (\cdots \oplus p^{A_{N}})\cdots)
\end{equation}
with similar expressions for the total momentum of the other body.
Involving the non linear addition in the definition of the total momenta  is motivated by the last remark in the previous section.
 In addition,
let us assume for simplicity that all the momenta of microscopic
constituents are identical, to wit $\forall_a\, p_{\mu}^{A_{a}}=
p_{\mu}^{A}$ etc. This  is a  simplistic model of a macroscopic
body, of course, but it makes it possible to capture the relevant
features of interacting macroscopical bodies at play in the soccer
ball issue.
 Then, once we chose to work in the connection normal coordinates,
we can use the fact that the non linear addition of colinear momenta is equal to the ordinary linear addition. Hence, in these coordinates we have that
 remarkably, $ \mathbf{P}^{A}_\mu$ equals just
$Np_{\mu}^{A}$. We can then easily sum up expressions (\ref{10}) over
$a$ to obtain
\begin{equation}\label{12}
    [ \mathbf{P}^{A} + \mathbf{P}^{B}]_{\mu} - [ {\tilde{\mathbf{P}}}^{A} +{\tilde{\mathbf{P}}}^{B}]_\mu =
\frac{\tilde{\Gamma}_\mu^{[\alpha \beta]}}{N\, M_P}
  \left(\mathbf{P}_{\alpha}^{A} \mathbf{P}_{\beta}^{B}- \tilde{\mathbf{P}}_{\alpha}^{A} \tilde{\mathbf{P}}_{\beta}^{B}\right)
\end{equation}

We thus arrive at the same conclusion as the naive argument (\ref{3naive}), that is the non linerarities for macroscopical bodies are damped by 
powers of $NM_{P}$ nstead of $M_{P}$,  but this time on firm ground.

We see therefore that in the limit of large number of elementary
constituents $N$ the linearly combined momenta satisfy with good
accuracy  the standard, linear conservation law. Indeed, if the
elementary microscopic constituents were atoms the ratio of the
second and first terms is of order of $10^{-18}$. Thus the soccer
ball problem is avoided. This is the main result of this note.

\subsection{Making the model slightly less idealized}

We can now also consider the case where the individual atomic
momenta fluctuate around the mean value $p$ thus \be p^{A_{a}}=
p^{A} + \delta p^{A_{a}} \ee where $\delta p^{A_{a}}$ are small
fluctuation that average to zero in time and when we sum over $a$.
We note that these fluctuations have to be small if the constituents
cohere into a  macroscopic body, as assumed. First one sees, using
the property of the connection normal coordinates,  and the fact
that $\sum_{a} \delta p^{A_{a}}=0$, that the non linear addition
(\ref{11}) differ from the linear one by a term equal to $
\frac1{M_{P}}\tilde\Gamma_{\mu}^{[\alpha\beta]} \sum_{a<b} \delta
p^{A_{a}}_{\alpha}\delta p^{A_{b}}_{\beta}$.  For a macroscopic
body, i.e. for large N,    one can safely estimate this term by
   examining the average value of    ${\tilde \Gamma}_\mu^{[\alpha \beta]}
     \delta p_\alpha^{Aa}\delta p_\alpha^{Ab}$,    also exploiting the fact that
   ${\tilde \Gamma}_\mu^{[\alpha \beta]}$ is antisymmetric.
   It is natural to assume that for large N
\begin{equation}\label{20}
\left< \delta p^{A_{a}}_{\alpha}\delta p^{A_{b}}_{\beta}\right> \sim
\frac{P^{2}}{N^{2}} \left(\delta^{ab} - \frac{v^av^b}N\right)
\sigma_{\alpha \beta}
\end{equation}
where $v^a$ is a vector with all components equal one, such that
$v^a\delta p_a=0$. The average $\langle \cdot \rangle$ taken here,
denotes an average over time, but
 a similar result is obtained for an ensemble average.
In light of eq.\ (\ref{20}) we can safely
estimate that the correction term is smaller than $\tilde\Gamma
PP/N^{3}M_{P}$ and therefore negligible or even vanishing since the fluctuation tensor
$\sigma_{\alpha\beta}$ is symmetric. These fluctuations also enter
the non linear conservation (\ref{10}) adding terms proportional to
$\tilde \Gamma_{\mu}^{\alpha \beta }\sum_{a} \langle \delta
p^{A_{a}}_{\alpha} \delta p^{B_{a}}_{\beta}\rangle
$ whose average vanish since the fluctuation of the body $A$ and $B$
are decorrelated.

\section{Conclusions}

We have seen that  the so called ``soccer ball problem''
is not present in the relative-locality framework
because the total momentum of a macroscopic body is equal, within
a very small margin, to the linear sum of momenta of the
constituents, i.e.
 \f\label{tot} \mathbf{P}^{total}=\sum_{a} p^{A_{a}}. \ff
We established this rigorously exploiting
 a choice of momentum-space coordinates for which this addition is linear
for colinear momenta. But we could turn around the reasoning and use
our argument  to show that the  quantity (\ref{tot})  is
macroscopically conserved in the interaction between two macroscopic
bodies. This suggest that  (\ref{tot}) can serve  as the definition
of what the momentum of a large composite system is. That is  we
could define the total momenta  as a quantity which is conserved in the
scattering processes provided it is also preserved by the internal
dynamics that bound the constituent of the macroscopical body
together.

Notice also that the number $N$ can be interpreted not only as a
number of elementary particles of the bodies, but as being
proportional to the number of elementary interactions. In the
realistic situation of scattering of two macroscopic bodies which
are approximately rigid not all the elementary constituents of
body $A$ interact with those of the body $B$. However,
each time a photon is emitted and absorbed by a  constituent of the
body the total momentum transfer is then quickly redistributed by the
internal interactions insuring the rigidity of the body to all the
constituents of the body
This redistribution happens if the initial
momentum transfer do not excite phonons interactions and can be
assume to happen over a time shorter than the total interaction
time between the two macroscopical bodies. This process involves at
least as many interactions as there are particles in the body which
makes the estimate (\ref{12}) valid even in this, more realistic
case.

We note that the idea that the soccer ball problem is solved because the non-linearities relevant for
the dynamics of a system composed of $N$ elementary particles
are suppressed by a mass scale $N M_P$ is not new. It has been
proposed a number of times before (see {\it e.g.}
Refs.~\cite{DSRB-2,EF-DSR, Jacobson}).
However, here we have shown it to be the case in a well defined class of theories
with curved momentum spaces.

We should also mention that there is an aspect of the soccer ball problem that deserved to be studied under the same lines developped here, which is to check consistency of the transformation properties
under boosts of the fundamental particles with the usual transformation properties of the total momenta. 

To conclude, we see that the soccer ball problem does not occur in
 theories satisfying the Principle of Relative Locality. In fact,
as stressed some time ago in \cite{Magueijo} the real
question is how to find a system which would exhibit large enough deformations  to be detectable in a feasible experimental setup.

\section*{ACKNOWLEDGEMENTS}

We thank Nima Arkani-Hamed for emphasizing the need to solve the
soccer ball problem and Michele Arzano, Florian Girelli, Sabine Hossenfelder, Etera
Livine, Joao Magueijo and Seth Major for discussions and criticism.
The work of J. Kowalski-Glikman was supported in part by grant
182/N-QGG/2008/0.
  Research at Perimeter Institute
for Theoretical Physics is supported in part by the Government of
Canada through NSERC and by the Province of Ontario through MRA.


\vskip -0.5cm


\begin{thebibliography}{99}


\bibitem{PRL}
  G.~Amelino-Camelia, L.~Freidel, J.~Kowalski-Glikman, L. Smolin,
  {\it The principle of relative locality},
  [arXiv:1101.0931 [hep-th]].

\bibitem{maggioNOsoccer} M.~Maggiore,
{\it The Atick–Witten free energy, closed tachyon condensation and deformed Poincar\'e symmetry},
Nucl.~Phys.~B647 (2002) 69.

\bibitem{sabiNOsoccer} S.~Hossenfelder,
{\it Multiparticle states in deformed special relativity},
Phys.~Rev.~D75 (2007) 105005

\bibitem{GRB-ll}L. Freidel and L. Smolin,{\it Gamma ray bursts probe the geometry of momentum space},  arXiv:1103.5626.

\bibitem{DSRB-2} J.~Magueijo and L. Smolin, {\it Generalized Lorentz invariance with an invariant energy scale}  Phys.~Rev.~D67 (2003) 044017 [arXiv:gr-qc/0207085]

\bibitem{EF-DSR}F. Girelli, E.  R. Livine, {\it Physics of Deformed Special Relativity: Relativity Principle revisited}, arXiv:gr-qc/0412004; {\it Physics of Deformed Special Relativity}, arXiv:gr-qc/0412079, Braz.~J.~Phys.~35 (2005) 432.



\bibitem{Jacobson}
  T.~Jacobson, S.~Liberati and D.~Mattingly,
  {\it Lorentz violation at high energy: concepts, phenomena and astrophysical
  constraints},
  Annals Phys.~321 (2006) 150
  [arXiv:astro-ph/0505267].




\bibitem{Magueijo}
  J.~Magueijo,
  ``Could quantum gravity be tested with high intensity lasers?,''
  Phys.\ Rev.\  D {\bf 73} (2006) 124020
  [arXiv:gr-qc/0603073].





\end{thebibliography}
\end{document}